# A compact superconducting nanowire memory element operated by nanowire cryotrons

Qing-Yuan Zhao[1], Emily A. Toomey[1], Brenden A. Butters[1], Adam N. McCaughan[2], Andrew E. Dane[1], Sae-Woo Nam[2], Karl K. Berggren[1]*

[1]Massachusetts Institute of Technology, Department of Electrical Engineering and Computer Science, Cambridge, MA, 02139, United States.

[2]National Institute of Standards and Technology, 325 Broad- way, Boulder, Colorado 80305, United States

*berggren@mit.edu

## Abstract:

A superconducting loop stores persistent current without any ohmic loss, making it an ideal platform for energy efficient memories. Conventional superconducting memories use an architecture based on Josephson junctions (JJs) and have demonstrated access times less than 10 ps and power dissipation as low as $10^{-19}$ J. However, their scalability has been slow to develop due to the challenges in reducing the dimensions of JJs and minimizing the area of the superconducting loops. In addition to the memory itself, complex readout circuits require additional JJs and inductors for coupling signals, increasing the overall area. Here, we have demonstrated a superconducting memory based solely on lithographic nanowires. The small dimensions of the nanowire ensure that the device can be fabricated in a dense area in multiple layers, while the high kinetic inductance makes the loop essentially independent of geometric inductance, allowing it to be scaled down without sacrificing performance. The memory is operated by a group of nanowire cryotrons patterned alongside the storage loop, enabling us to reduce the entire memory cell to 3 μm × 7 μm in our proof-of-concept device. In this work we present the operation principles of a superconducting nanowire memory (nMem) and characterize its bit error rate, speed, and power dissipation.

## 1. Introduction:

A fast, energy efficient, and scalable memory is an essential component in building a computer. Superconducting digital circuits based on single flux quantum (SFQ) logic offer fast calculation speed and low power dissipation, motivating the development of a superconducting computer for supercomputers and big data centers[1]. Basic logic gates, analog-to-digital converters, and small processors have been demonstrated by SFQ circuits. However, the challenge of creating a high speed, low power, and scalable memory that operates at cryogenic temperatures for SFQ compatibility remains an obstacle to the development of a practical superconducting computer. Several technologies in the past have been built to achieve this goal. One approach involves using a hybrid architecture that combines SFQ circuits and CMOS memories[2]. Scaling up CMOS memory to the level of a superconducting computer is relatively easy, benefiting from technologies developed in the advanced semiconductor industry. But, as the CMOS circuit requires large voltage input, the amplification interface between SFQ and CMOS units consumes the majority of the power dissipation and limits the operating speed. Another hybrid approach uses multiple layers of magnetic materials to create a superconducting-ferromagnetic-superconducting (SFS) junction[3,4]; however, this technique demands careful tuning of the materials to enable a scalable array at cryogenic temperature.

Compared to these hybrid approaches, a technique relying on memories and readout circuits made entirely of superconductors may be more straightforward, as they share the same signal levels, temperature dependences, and fabrication processes as SFQ circuits. A conventional all-superconductor memory stores bit information in a superconducting loop and uses SFQ circuits to enable addressing, writing and reading operations[5]. However, the development of a scalable Josephson junction (JJ)-based memory has been slow due to several limitations[6]. First, reducing the area of a JJ below 1 $\mu m^2$ makes it increasingly difficult to fabricate a junction of high current density and a high yield. In addition, the superconducting loop requires an inductor of at least few pH to ensure the conditions for single flux quantum operations, increasing the overall area required by the device. The total area is further increased by the transformers and SQUID amplifiers used in addressing, writing and reading operations of the storage elements. Furthermore, since magnetic coupling is typically used in SFQ circuits, adjacent memory cells must be far enough apart to avoid crosstalk, limiting the density of memory arrays.

Here, we demonstrate an alternative superconducting memory made entirely of lithographic nanowires (nMem). We use superconducting nanowire devices, which are patterned together with the nanowire storage loop in a very compact size, to enable operations for addressing, writing and reading. In comparison to Josephson based memory elements, the nMem offers multiple advantages. The minimum feature size defined as the width of a nanowire is typically ~100 nm, smaller than a Josephson junction by 1~2 orders of magnitude. The entire memory cell is patterned from a single thin (~7 nm) film and could be patterned in multiple layers for an even higher scalability, making it promising for large arrays. Additionally, the kinetic inductance of a thin nanowire is about two orders of magnitude larger than its geometric inductance, allowing the superconducting loop to be scaled down while maintaining the high inductance required for storage. Furthermore, because magnetic fields penetrate through the thin nanowires, the nMem is not sensitive to perturbation by magnetic fields and thus may be densely packed into an array without crosstalk. Using the kinetic inductance to shrink the size of a superconducting loop was demonstrated in Ref.[7]. The authors designed a superconducting loop into a nanoSQUID and operated it as a memory by sending current pulses or applying magnetic field. In this work, the memory combines with on-chip cryotron devices, which are used for addressing, writing and reading the storing loop. Therefore, we can fully operate the memory with digital pulses and characterize its bit error rate. We have also previously demonstrated that SFQ pulses can trigger a nanowire cryotron [8], suggesting that nMems can be integrated with RSFQ circuits through an interface circuit made from cryotrons.

## 2. Memory operation:

Figure 1a shows the schematic diagram of a single-memory cell. A superconducting loop stores the bit information in the form of a persistent current, while a thermally coupled cryotron, which we refer to as heat-Tron (hTron), enables the write operation and a current-crowding cryotron (yTron) reads the stored persistence current nondestructively. The loop and the cryotrons were patterned together within a 3 μm × 7 μm area as shown in Fig.1b. Figure 1c shows experimental waveforms for writing and reading the two different states. The detailed operation principle will be discussed following the descriptions of the cryotron devices.

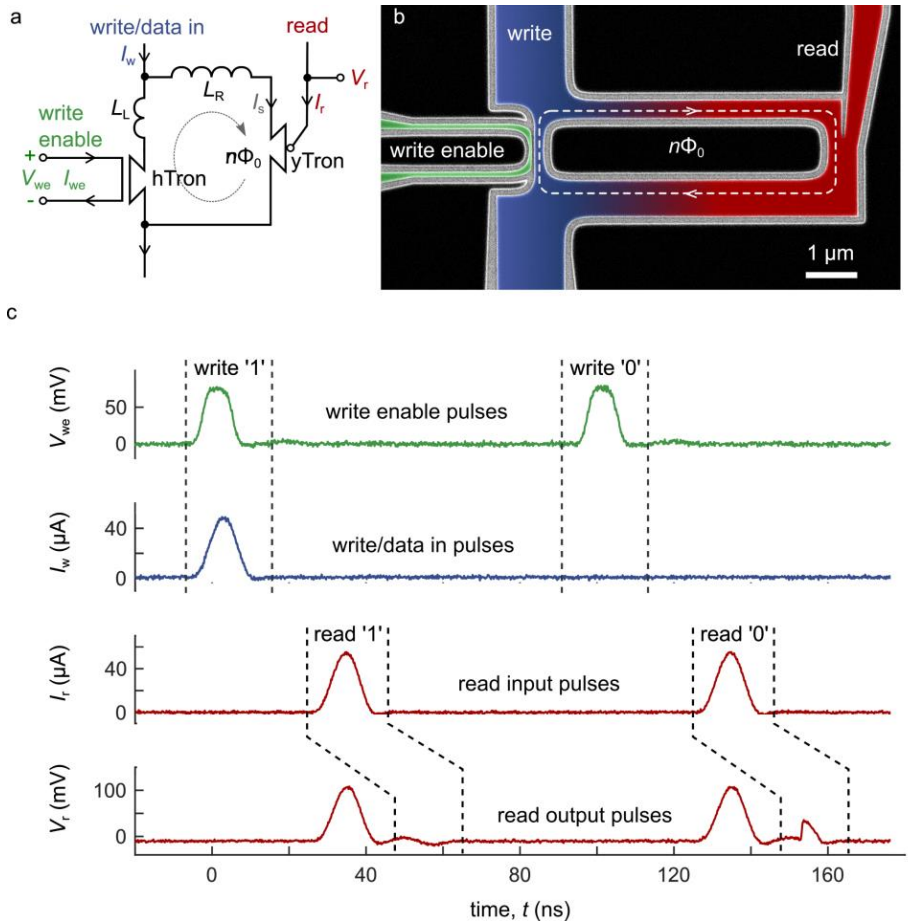

**Figure 1 | A superconducting nanowire memory operated by nanowire cryotrons.** (a) Circuit schematic diagram of a single memory pixel. Three ports (write-enable, write, and read) are used for operating the memory. $L_L$ and $L_R$ are the inductances of the left and right nanowires that form the superconducting loop. Taking a calculated kinetic inductance of 60 pH per square, the values of $L_L$ and $L_R$ are $L_L = 0.37$ nH and $L_R = 1.37$ nH. (b) SEM image of the nMem. The black area is the niobium nitride film, while the white area is the substrate (Si substrate with a thermal oxidized surface). (c) Experimental pulses for writing and reading bits '1' and '0'. To read the memory, we used the same input port for sending the bias and reading the output. In the output pulse $V_r$, the first pulse is the leakage signal from the biasing pulse $I_r$, while there

will be a second pulse appears after the leakage pulse only if the storage state is '0'. The circuit diagram is shown in Fig. 6a.

## 2.1 hTron characterization

A large memory requires bit selection scheme to operate either an individual bit or a group of bits, i.e., a word. In the nMem, the superconducting loop stores the bit information while the cryotrons work for the bit selection. We use the heat-Tron (hTron) as a selection line to enable the write operation. Only when the hTron is triggered, can bit information be written into the superconducting loop. Since heat is generated during the operations of the hTron, it is important to characterize the power dissipation and switching speed of the hTron, which could limit the overall performance of the memory element.

The hTron is a nanowire cryotron device comprised of two isolated nanowires placed close together with a typical spacing of 40 nm. We refer to the narrower nanowire as the gate and the wider nanowire as the channel. As shown in Fig. 1b, an hTron is on the left side of the memory with its channel forming part of the storage loop. When an input pulse switches the hTron gate from the superconducting state to the resistive state, the gate dissipates power and increases the local temperature of the channel through Joule heating, suppressing its critical current. Applying a biasing current to the channel greater than the suppressed critical current will cause the channel nanowire to switch. In this way, the switching of the hTron channel nanowire dictates the opening of the superconducting loop for fluxons to enter (write '1') or exit (write '0'). The electrical isolation between the hTron gate and channel minimizes crosstalk between the port for selecting a memory loop and the ports for writing and reading the loop, which is a promising feature for a multiplexing memory array.

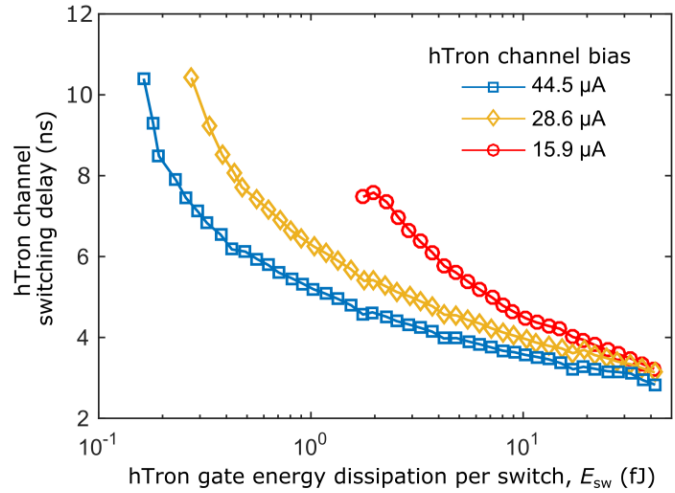

**Figure 2 | Tradeoff between the delays for switching the hTron channel and energy dissipations on the hTron gate.** This measurement was performed with sweeping the biasing current to the hTron channel, as indicated in the legend. The data plotted were collected from the switching cases that the channel always switched by a gate pulse. We observed that the channel switched probabilistically if $E_{sw}$ was too weak.

We characterized an individual hTron device isolated from the storage loop. We found that there was a tradeoff between the dissipation power on the gate and the delay for switching the channel. To observe this effect, we sent fast pulses to the hTron gate and channel to measure the delay between the input pulse to the hTron gate and the switching time of the channel. Delays of the cables and amplifiers were removed after calibration. The width of the pulse sent to the hTron gate was fixed at $\tau_p = 8$ ns, while the high level of the pulse was swept in order to generate different energy dissipations. We assumed that the current through the gate wire was held constantly at a self-heating current of $I_{hg} = 2$ μA and that all of the input voltage was applied on the gate. Thus, the energy dissipation per switch on the gate was calculated by using $E_{sw} = V_{gh} \times I_{hg} \times \tau_p$. The data in Fig.2 shows that the switching delay is a function of the biasing current on the hTron channel and the energy dissipation on the gate. It took longer for the channel to switch when less energy was dissipated on the gate and when the channel was biased at a lower current.

Compared to SFQ circuits, the hTron is more energetically expensive per switch and requires a longer time for completing a write operation. This would limit the application of the hTron in a fast and energy efficient memory array or a logic circuit. A future multilayer design that stacks the hTron gate on top of channel with a thin insulting layer would enhance the thermal coupling, making the hTron faster and more energy efficient. In this work, however, the electrical isolation between the hTron gate and channel makes it a valuable tool for characterizing the memory operations.

## 2.2 yTron characterizations

To read the stored bit information, i.e. the circulating current, there are a destructive approach and a non-destructive approach. The destructive readout approach can be done by sensing the switching current of the memory loop through the write port. We will discuss it in section 3.3. The non-destructive readout approach uses the current-crowding cryotron (yTron), which senses the circulating current of the superconducting loop through the read port. As the detection happens in the read nanowire, the superconducting state of the storage loop is maintained, enabling us to read the stored bit for multiple times. We would like to discuss the operation principles and characterize the sensitivity of the yTron in advance for better presenting the memory results as following.

The yTron is a device made from two nanowires joined together with a sharp intersection point. It uses the current-crowding effect to control the switching current of one arm with the bias current through the other[9]. Details of the operation principle of a yTron are described in Ref. [10]. Here, we will discuss the readout approach of a memory with an integrated yTron. The information stored in the superconducting nanowire memory is in terms of the number of fluxons. The trapped fluxons $n\Phi_0$ generate a persistent current of $n\Phi_0/L_L$, where $L_L$ is the total loop inductance and $\Phi_0$ is the magnetic flux quantum. This persistent current is also a biasing current to one arm (sensing arm) of the yTron device, and thus controls the switching current of the other arm (detecting arm) of the yTron. Therefore, we can read the different fluxons stored in each state by measuring the difference of the switching current of the yTron detecting arm. One promising feature of using the yTron as a readout tool is that reading the detecting arm has no effect on the superconducting state of the sensing arm attached to the storing loop. Therefore, the read operations are nondestructive.

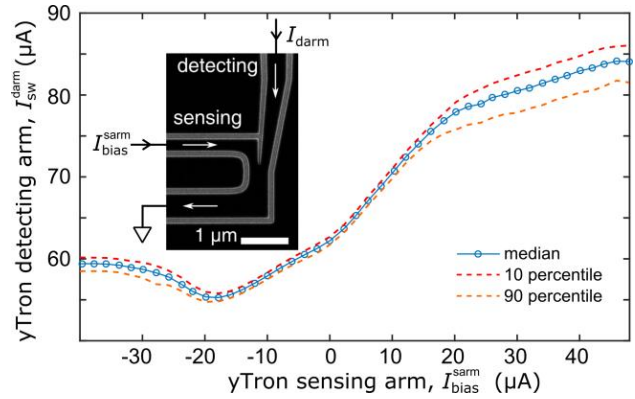

**Figure 3 | Sensitivity and operation margin of an isolated yTron.** The yTron is designed with the same dimensions as the one used in the memory (300 nm wide for both arms). At each level of the bias current sent to the sensing arm, we measure the statistics of the switching current of the yTron detecting arm with 1,000 sweeps. The blue trace shows the median value of the detecting arm switching current, while the 10 and 90 percentile values are shown as the dashed lines.

We measured the dependence of switching current of the detecting arm $I_{sw}^{darm}$ to the biasing current through the sensing arm $I_{bias}^{sarm}$ of a separate yTron which had the same geometry as the one used in the memory. The sensitivity of the yTron is defined as the derivative $d(I_{sw}^{darm})/d(I_{bias}^{sarm})$, which is the slope of the $I_{sw}^{darm}$ vs. $I_{bias}^{sarm}$ curve as shown in Fig. 3. We observed that the yTron responded to the change over a wide range of $I_{bias}^{sarm}$ but with varied sensitivity. The highest and most constant sensitivity (~0.8) was over the range 0 µA $< I_{bias}^{sarm} <$ 20 µA, which was where we operated the persistent currents in the memory.

### 2.3 Memory operation diagram

With knowledge of the cryotron devices, we can now discuss the operations of the memory shown in Fig. 1. To write currents into the storage loop, representing bit '1', we sent a pulse through the write port to bias the wires in the memory loop below the level that the loop can switch. Afterwards, another pulse was sent through the gate nanowire of the hTron, representing the write-enable port. The write-enable pulse then switched the hTron channel, allowing about 15 fluxons to enter the loop. To write a lower current into the loop, representing bit '0', we only sent a write-enable pulse without biasing the wires in the memory loop; this switched the hTron gate in order to either erase the '1' state if it had been written by the previous operation or maintain the '0' state. The read operation was performed by reading voltage pulses generated from the yTron detecting arm. As the stored currents for states '1' and '0' determined two different switching currents ($I_{sw1}^{darm} > I_{sw0}^{darm}$), we sent a pulse to bias the yTron's detecting arm to a current level close to ($I_{sw0}^{darm}+I_{sw1}^{darm}$)/2. Therefore, if the memory state was '1', we read no pulse from the yTron's detecting arm. Otherwise, if the memory state was '0', the yTron's detecting arm switched and a voltage pulse was observed.

We simulated the memory circuit to understand how the currents in the memory loop changed during a writing '1' operation. In particular, we studied how current pulse from the write port ($I_w$) split between the left arm ($I_{left}$, through the hTron channel) and the right arm ($I_{right}$, through the yTron sensing arm) of the storing loop. As shown in Fig.4a-c, before the hTron was turned on, $I_w$ split to $I_{left}$ and $I_{right}$ with a ratio α/(1-α) = $L_{right}/L_{left}$, where $L_{right}$ and $L_{left}$ were the inductances of the right and left sides of the nanowire loop, respectively. When $I_w$ reached the highest level $I_w^{high}$ and then the hTron was turned on, the switching current of the left arm $I_{sw}^{left}$ was suppressed below $α×I_w^{high}$. Thus, the left wire switched into resistive state, expelling the bias current to the right wire. Diversion of bias current reduced $I_{left}$ to a level $I_{res}$ at which the resistive state in the left wire could no longer be maintained, allowing it to return to superconducting state. After the hTron was off and $I_w$ was removed, $I_{left}$ and $I_{right}$ reduced following the same splitting ratio $L_{right}/L_{left}$. When all of the input pulses were removed, the superconducting loop stored a circulating current $I_{store} = αI_w^{high} – I_{res}$.

The simulation indicates that a higher $I_{store}$ will be written into the storing loop for a higher $I_w^{high}$. However, too much input current will switch both arms when $I_w^{high} - I_{res} > I_{sw}^{right}$, where $I_{sw}^{right}$ is the switching current of the right wire. We observed a sharp transition when $I_{sw}^{right}$ was too high as shown in Fig.4d. We measured the switching current of the yTron's detecting arm $I_{sw}^{darm}$, which was proportional to $I_{store}$, at different levels of $I_w^{high}$. Increasing $I_w^{high}$ increased $I_{sw}^{darm}$ until $I_w^{high}$ = 48 µA, agreeing with our simulation results. As we showed in Fig. 2, to make the hTron switched deterministically, the write-enable pulse had to be enough powerful to switch the superconducting loop. We found the linear increase shown in Fig. 4d started at a higher $I_w^{high}$ for a weaker write-enable pulse, which agreed with our previous data of an individual hTron shown in Fig.2.

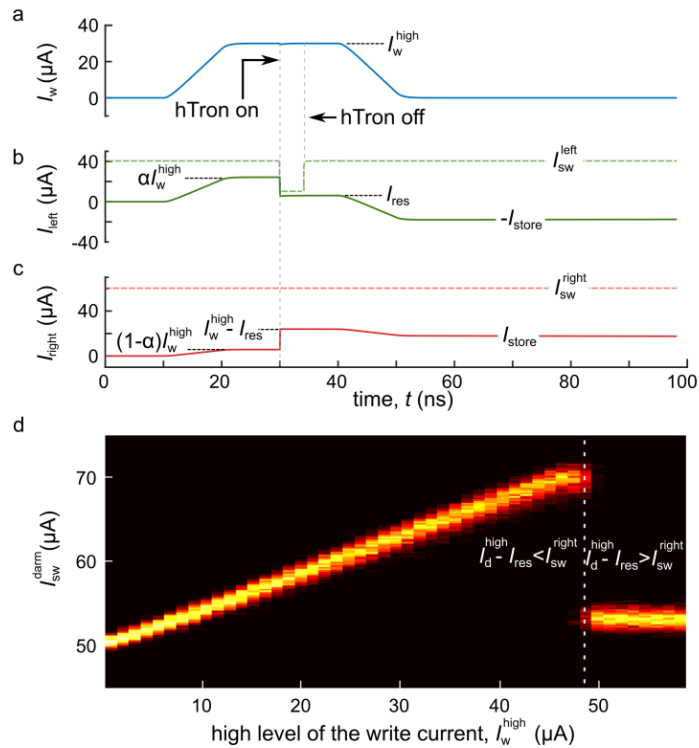

**Figure 4 | Writing a persistent current in the memory** (a-c) SPICE simulation of writing a persistence current $I_{store}$ (writing bit '1') into the memory. A square pulse was sent into the loop from the write port. The inductance of the two arms of the loop split the input current to the left ($I_{left}$) and right ($I_{right}$) arms. The dashed lines indicate the switching currents for both arms. At time $t$ = 30 ns, the hTron was turned on to suppress the switching current of the left nanowire, triggering the left nanowire to switch into the normal state and increasing the current of the right nanowire. When the input pulse was removed, a persistent

current $I_{store}$ was stored in the loop. (d) Experimental data of the dependence of $I_{sw}^{darm}$ on $I_{w}^{high}$. The color shows the probability of the switching current at each $I_{w}^{high}$. The maximum $I_{sw}^{darm}$ occurred when $I_{w}^{high}$ was ~48 µA, above which both wires of the storing loop switched into the normal state.

## 3. Memory Characterizations

### 3.1. Bit Error Rate

To ensure correct write and read operations, the nanowire memory must perform with a very low bit error rate (BER). As we discussed in previous sections, the write operation can be deterministic if we dissipated enough energy on the hTron gate and set a proper value for $I_{w}^{high}$. The bias margin of the write operation could be much wider than the bias margin of the read operations, if energy efficiency was not seriously considered. Here, we focus on errors caused by the read operations. In specifically, we would like to characterize the bias margin of the yTron for ensuring the memory operations of an acceptable BER. To ensure the write operations of no errors, the writing current was fixed at $I_{w}^{high}$ = 32 µA and the energy dissipation of the hTron pulse was set to 13 fJ (pulse width was 8 ns, pulse high level was 0.8 V).

The BER measurements made on our devices are shown in Fig. 5. In every measurement cycle, we first wrote a random bit '1' or '0' to the nMem. Afterwards, we sent a pulse with fixed amplitude to the yTron's detecting arm to read the memory state. If bit '1' was stored, the yTron was expected to be in superconducting state and no output voltage pulse would be detected. In the opposite case when bit '0' was written, we expected to measure a voltage pulse. Because the operation signals for the nMem were pulses, we first generated a pseudorandom binary sequence (PRBS), and then used this sequence to trigger a second arbitrary waveform generator (AWG) to produce pulses of fixed width and amplitude. As the rising-edge triggering mode was used, the output pulse train only indicated the transitions from bit '0' to bit '1'. On average, one quarter of the PRBS bits produced a pulse for writing '1'. The timing diagram of the operation patterns are illustrated in Fig. 5b.

We used a counter to record the total number of operations $N_{tot}$, the number of bit '1' writes $N_{W1}$, and the number of bit '0' reads $N_{R0}$, from which the BER can be calculated by BER = $1-N_{R0}/(N_{tot}-N_{W1})$. The maximum $N_{tot}$ was set to be $3 \times 10^7$, giving a lowest measurable BER of $4.4 \times 10^{-8}$. As shown in Fig. 5c, when the read pulse level was too far below the switching current of the detecting arm, the yTron did not always switch when bit '0' was written, causing the W0R1 errors (write bit '0' but read bit '1'). When the reading pulse was too high, the yTron detecting arm switched even when we wrote bit '1', causing the W1R0 errors (write bit '1' but read bit '0'). Only when the reading pulse was in an optimal range could correct operations be obtained. The read operation margin was defined as the biasing range at a fixed BER. For a BER on the order of $10^{-7}$, the biasing margin was from 52.4 µA to 57.0 µA. Fits to the trench of the BER curves suggested that a BER lower than $10^{-10}$ could be achieved but with a narrowed operation margin.

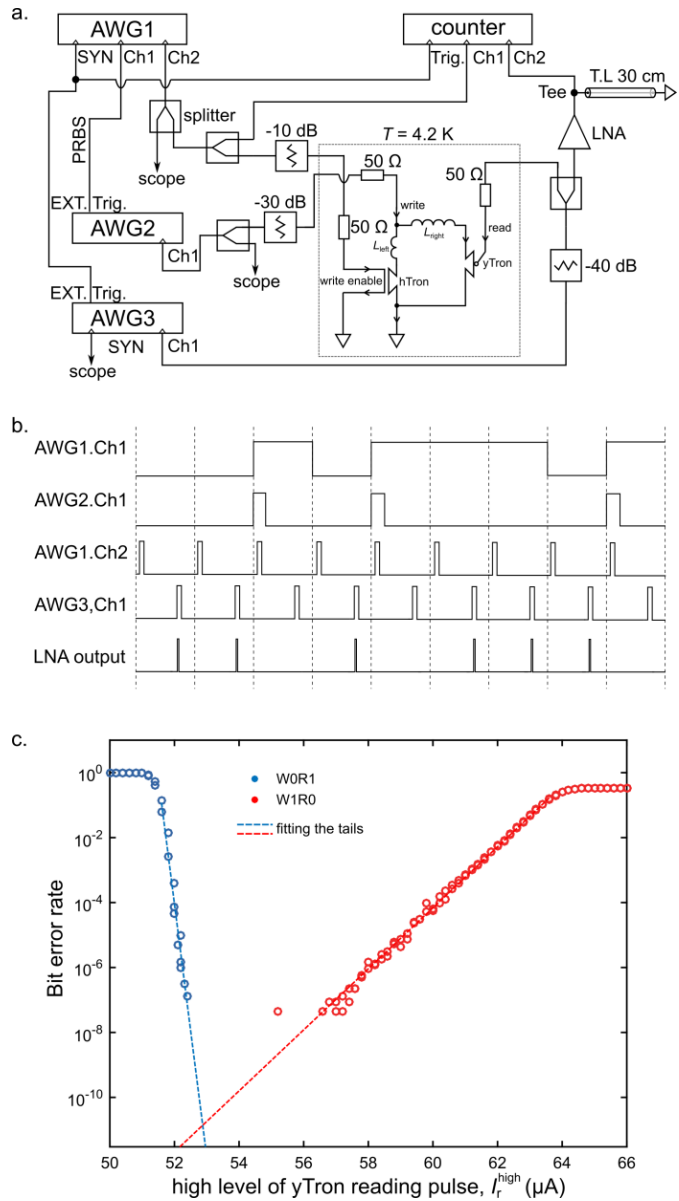

**Figure 5 | Bit error rate measurement of the nMem.** (a) Measurement setup for generating random writing pulses to the memory and recording the BER. The pulses were prepared and attenuated at room temperature. A short-terminated coaxial cable was connected at the amplifier's output port to extract the leading edge of the voltage pulses from the yTron detecting arm. (b) Timing diagram of the operation pulses. (c) Measured BER at different reading currents. The testing resolution, determined by the maximum number of writing pulses, gave a lowest measurable BER of $4.4 \times 10^{-8}$. The dashed lines are fitting lines from the measured data, indicating a possible BER less than $10^{-10}$.

## 3.2. Non-destructive readout

In the reading operations, although the yTron's detection arm switches to normal state and produces a voltage pulse, the superconducting state of the storing loop is not disturbed. In this way, the yTron offers a non-destructive readout of the nMem. To demonstrate non-destructive readout by the yTron, we wrote to the memory once, but read its state multiple times. Unlike the read operations used in measuring the BER, we sent ramped pulses of an amplitude higher than the maximum switching current to the yTron detecting arm to determine when it switched, from which we calculated the switching currents of the yTron's detecting arm at different memory states. As a result, we forced the yTron to switch for reading both bit '1' and bit '0'.

As shown in Fig.6a, 400 read operations were executed within 200 µs after one write operation. When bit '1' was written, the yTron's detecting arm switched later along the bias current ramp, indicating that it had a higher switching current. In the case of writing bit '0', the yTron's detecting arm switched earlier. The measured switching delays in both cases decayed over time, presumably because the local temperature was increased by the heat generated from the yTron's detecting arm when it switched to the normal state. When fewer reading pulses were sent to the nMem within a longer time frame, the temperature had a longer time to cool down, resulting in reading of stable switching currents of the yTron's detecting arm. As shown in Fig. 6b, we performed 180 read operations within 900 µs. The measured switching currents were more stable than the data shown in Fig. 6a.

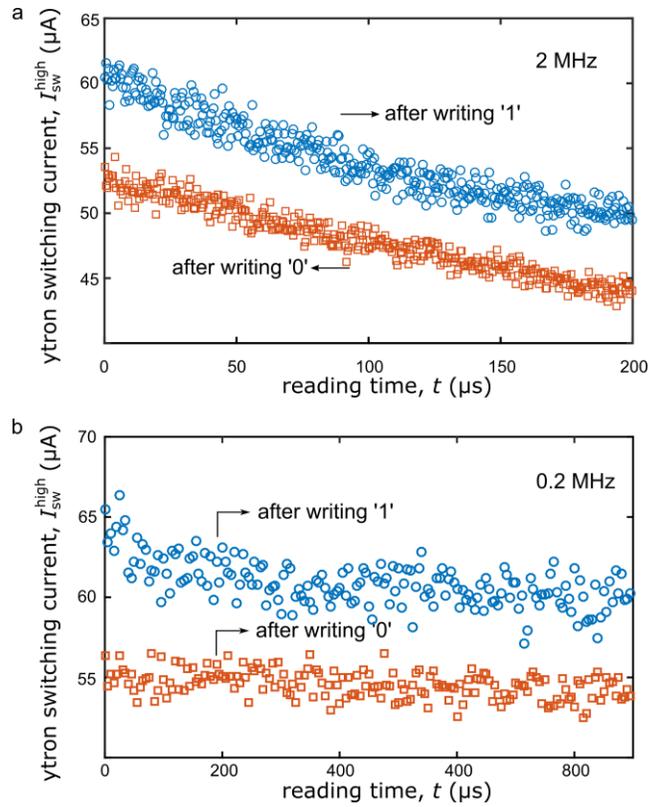

**Figure 6 | Non-destructive readout of the yTron** Non-destructive readout of the nMem via the yTron at repetition rates of 2 MHz (a) and 0.2 MHz (b). The yTron input reading pulse increased from 0 μA to 70 μA within a 100 ns rise time. The pulse duration was 200 ns wide. We calculated the switching currents of the yTron's detecting arm from its switching delay with respect to the input pulse. Red dots represent measurements for reading bit '1' ($I_w^{high}$ = 32 μA) and blue dots represent reading bit '0' ($I_w^{high}$ = 0 μA).

### 3.3. Bipolar operation without hTron

In the present nMem design, the hTron dominates the overall energy consumption. As mentioned previously, a stacked hTron design with the gate nanowire on top of the channel would likely increase thermal coupling and reduce the power dissipation; however, this tactic requires the development of a multilayer process. An alternative way to reduce energy costs would be to avoid using the hTron by writing to the nMem with bipolar pulses through the write/data-in port. We demonstrated this bipolar operation using the same nMem device while leaving the hTron gate grounded. As shown in Fig. 7a, for each write operation, we always sent a negative pulse of higher amplitude through the write/data-in port to force the memory loop to switch regardless of its previous state, acting as a clear operation. This also generated a level of the persistent current representing bit '0'. To write bit '1', a positive pulse was sent after the negative pulse. The amplitude of the positive pulse was adjusted to a level such that only the left arm of the memory loop switched while the right arm remained superconducting, which was equivalent to the previous operation of writing bit '1' when the hTron was used.

In the bipolar operation, the negative pulse switched both arms of the loop into resistive state. Therefore, the resulted dissipation was proportional to the duration of the high level of the input pulse. The same situation occurred when we read the state by switching the yTron. For these operations that created a long voltage state, the power dissipation could be reduced by applying shorter pulses. For writing bit '1' into the storage loop, we chose a proper level of the positive $I_w$ pulse. Thus, only the left nanowire of the storage loop switched into normal, while the right nanowire kept into superconducting state, acting as an inductive shunt. This resulted in a short duration of the resistive state and a weak output voltage pulse that we cannot measure directly. To estimate the power dissipation for this switching dynamic, we performed a SPICE simulation which included the electrothermal dynamics of the superconducting nanowire. As shown in Fig.7b, switching the left wire produced a hotspot resistance which grew to a maximum value of ~150 Ω after ~40 ps. While in the resistive state, the switched wire dissipated power as shown in Fig.7c. During the lifetime of the resistive state, the total energy dissipated was $2\times10^{-18}$ J. This energy consumption is significantly lower than the experimentally evaluated write operations using the hTron and is close to what a Josephson junction costs in an SFQ circuit, which has a typical value of $10^{-19}$ J. This simulation indicates that, if we could control the lifetime of the resistive state of a switched nanowire by inductive or resistive shunting for the clear and read operations, the overall power dissipation could be greatly reduced.

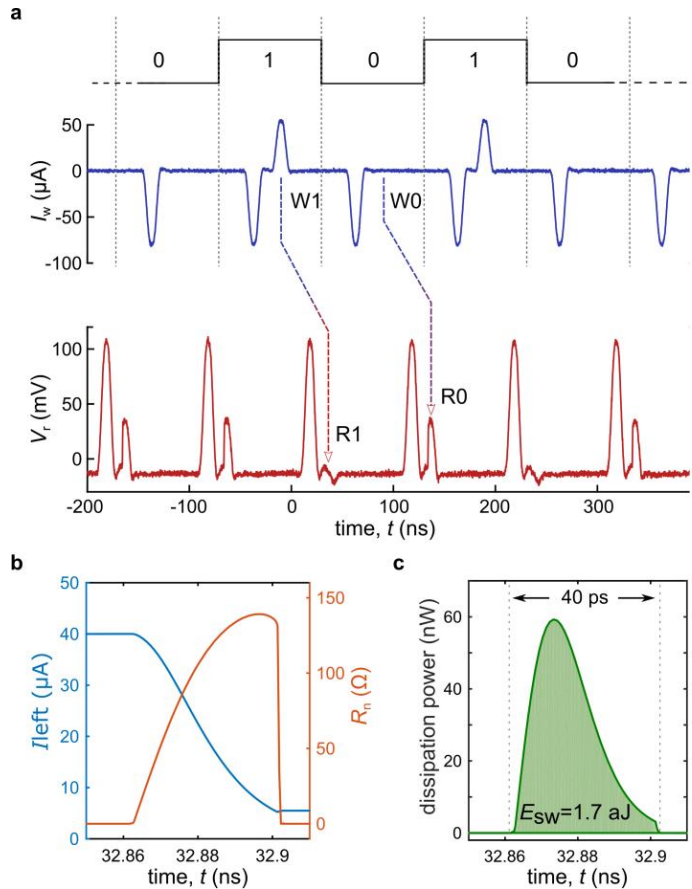

**Figure 7 | Bipolar operations of writing states to the memory without using an hTron** Experimental bipolar pulses to write bit '0' and '1' to the nMem. The stored states were read by the yTron. (b) SPICE simulation of the switching dynamics of the nMem when bit '1' was written. The switching current of the left nanowire was set to 40 µA. (c) Power dissipation calculated from the SPICE simulation of the current and voltage through the switched nanowire. The memory dissipated a total energy of $1.7 \times 10^{-18}$ J for a lifetime of normal state of 40 ps.

## 4. Conclusion

In this work, we have demonstrated a superconducting memory made entirely from nanowire devices fabricated together on a single plane. We discussed the advantages of the nMem and described its operation principles. The nMem has a compact size which is promising for scaling up to a large memory array; while our proof-of-concept device was 3 µm x 7 µm, the nMem can be minimized in future iterations by reducing the nanowire width and loop dimensions while maintaining a high kinetic inductance. Multilayer fabrication may also allow for arrays of even higher density. We measured a minimum BER less than $10^{-7}$, indicating that the memory is reliable. The nMem was operated in the electrothermal regime, where a normal resistance needed to sustain to enable write and read operations. This operation regime could be analogous to a JJ operated in a latched mode. Due to heating from the normal resistance, the performance metrics of speed and power dissipation were not competitive to the performance of JJs operated in flux regime, i.e. the JJ memories in SFQ circuits. To speed up the memory operations and reduce the power dissipation of an nMem, it may be possible to operate the nanowire in flux regime by resistively shunting to suppress Joule heating during switching [11][12]. Therefore, we envision that the nMem's performance could eventually match the speed and power dissipation of RSFQ circuits.

## Acknowledgements

The authors thank James Daley and Mark Mondol for their technical support in nanofabrication, and Ling-dong Kong in preparing the figures. This research was supported by the Office of the Director of National Intelligence (ODNI), Intelligence Advanced Research Projects Activity (IARPA), via contract W911NF-14-C0089. Andrew Dane was supported by NASA Space Technology Research Fellowship (award number NNX14AL48H). Emily Toomey was supported by the National Science Foundation Graduate Research Fellowship Program (NSF GRFP) under Grant No. 1122374.

## References:

[1]     Holmes D S, Member S, Ripple A L and Manheimer M A 2013 Energy-Efficient Superconducting Computing — Power Budgets and Requirements *IEEE Trans. Appl. Supercond.* **23** 1701610–1701610

[2]     Van Duzer T, Lizhen Zheng, Whiteley S R, Kim H, Jaewoo Kim, Xiaofan Meng and Ortlepp T 2013 64-kb Hybrid Josephson-CMOS 4 Kelvin RAM With 400 ps Access Time and 12 mW Read Power *IEEE Trans. Appl. Supercond.* **23** 1700504–1700504

[3]     Gingrich E C, Niedzielski B M, Glick J A, Wang Y, Miller D L, Loloee R, Pratt Jr W P, Birge N O, Pratt R L W P and Birge N O 2016 Controllable 0-pi Josephson junctions containing a ferromagnetic spin valve *Nat. Phys.* **online** 1–6

[4]     Baek B, Rippard W H, Benz S P, Russek S E and Dresselhaus P D 2014 Hybrid superconducting-magnetic memory device using competing order parameters. *Nat. Commun.* **5** 3888

[5]     Nagasawa S, Hashimoto Y, Numata H and Tahara S 1995 A 380 ps, 9.5 mW Josephson 4-Kbit RAM Operated at a High Bit Yield **5**

[6]     Tolpygo S K 2016 Superconductor digital electronics: Scalability and energy efficiency issues *Low Temp. Phys.* **42** 361–79

[7]     Murphy A, Averin D V and Bezryadin A 2017 Nanoscale superconducting memory based on the kinetic inductance of asymmetric nanowire loops *New J. Phys.* **19** 63015

[8]     Zhao Q-Y, McCaughan A N, Dane A E, Berggren K K and Ortlepp T 2017 A nanoCryotron comparator can connect single-flux quantum circuits to conventional electronics *Supercond. Sci. Technol.* **30** 1–17

[9]     Clem J R and Berggren K K 2011 Geometry-dependent critical currents in superconducting nanocircuits *Phys. Rev. B* **84** 174510

[10]    McCaughan A N, Abebe N S, Zhao Q-Y and Berggren K K 2016 Using Geometry To Sense Current *Nano Lett.* **16** 7626–31

[11]    Brenner M W, Roy D, Shah N and Bezryadin A 2012 Dynamics of superconducting nanowires shunted with an external resistor *Phys. Rev. B* **85** 224507

[12]    Toomey E, Zhao Q-Y, McCaughan A N and Berggren K K 2017 Frequency pulling and mixing of relaxation oscillations in superconducting nanowires, arxiv